\documentclass[10pt,a4paper]{article}
\usepackage[utf8]{inputenc}
\usepackage{amsmath}
\usepackage{amsfonts}
\usepackage{amssymb}
\usepackage{graphicx}
 \usepackage{float}
\usepackage{indentfirst}
\usepackage{authblk}

\usepackage{bm}
\usepackage{xcolor} 
\usepackage{soul}   % strikethrought text

\title{Acceleration of the NVT-flash calculation for multicomponent mixtures using deep neural network models}
\author[1]{Yiteng Li}
\author[1]{Tao Zhang}
\author[1,2]{Shuyu Sun\thanks{Corresponding author: shuyu.sun@kaust.edu.sa}}
\affil[1]{
	Physical Science and Engineering Division (PSE), 
	King Abdullah University of Science and Technology (KAUST), 
	Thuwal, 23955-6900, Saudi Arabia}
\affil[2]{
	Institute of Geophysics and Geomatics, 
	China University of Geosciences, Wuhan 430074, P.R. China}
\date{}

\begin{document}

\maketitle % keep this until we decide to which journal we submit our manuscript. Then the author information will be complemented.
	
\section*{Abstract}

Phase equilibrium calculation, also known as flash calculation, has been extensively applied in petroleum engineering, not only as a standalone application for separation process but also an integral component of compositional reservoir simulation. Previous researches devoted numerous efforts to improving the accuracy of phase equilibrium calculations, which place more importance on safety than speed. However, the equation-of-state based flash calculation consumes an enormous amount of computational time in compositional simulation and thus becomes a bottleneck to the broad application of compositional simulators. Therefore, it is of vital importance to accelerate flash calculation without much compromise in accuracy and reliability, turning it into an active research topic in the last two decades. 

With the rapid development of computational techniques, machine learning brings another wave of technology innovation. As a subfield of machine learning, the deep neural network becomes a promising computational technique due to its great capacity to deal with complicated nonlinear functions, which attracts increasing attention from the academia and industry. In this study, we establish a deep neural network model to approximate the iterative NVT-flash calculation. A dynamic model designed for NVT flash problems is iteratively solved to produce data for training the neural network. In order to test the model's capacity to handle complex fluid mixtures, three real reservoir fluids are investigated, including one Bakken oil and two Eagle Ford oils. Compared to previous studies that follow the conventional flash framework in which stability testing precedes phase splitting calculation, we incorporate stability test and phase split calculation together and accomplish both two steps by a single deep learning model. The trained model is able to identify the single vapor, single liquid and vapor-liquid state under the subcritical region of the investigated fluids. A number of examples are presented to show the accuracy and efficiency of the proposed deep neural network. It is found that the trained model makes predictions at most 244 times faster than the iterative flash calculation under the given cases. Even though training a multi-level network model does take a large amount of time that is comparable to the computational time of flash calculations, the one-time offline training process gives the deep learning model great potential to speed up compositional reservoir simulation. \\

\textbf{Keywords}: Deep learning, NVT-flash calculation, Dynamic model and realistic hydrocarbon mixture % if desired

\section{Introduction}

% 1st paragraph
Phase equilibrium calculations are not only performed independently, for instance, in separation process, but also play a significant role in compositional simulation during which complex phase behaviors often take place and accompany with considerable mass transfer between hydrocarbon phases. As the secondary recovery method fails to bring further economical production from conventional reservoirs, enhanced oil recovery (EOR) techniques, like miscible flooding, are increasingly employed to improve oil productivity. The simulation of EOR processes requires accurate prediction of phase behavior that could be modeled by the commonly-used equation of state (EOS), such as Peng-Robinson EOS \cite{Peng1976} or Soave-Redlich-Kwong EOS \cite{Soave1972}, thus placing great importance on the EOS-based phase equilibrium calculation. However, due to the large nonlinearity of phase equilibria problems, the popular iterative solution method for stability testing and phase behavior calculation at given conditions (usually pressure, temperature and feed composition) is computationally expensive. It has been reported that up to 70\% of computational time is consumed by flash calculation in compositional reservoir simulation. Although good initial approximations for phase splitting calculation can be obtained from stability test, the accuracy is ensured at the cost of efficiency, which is attributed to the repeated stability analysis, as well as the high-dimension formulation and iterative algorithm itself. Since speed is one of critical concerns in current compositional simulator, speeding up flash calculation without too much compromise in accuracy and reliability is still an active research topic in both academia and industry.

% 2nd paragraph
Compared to the black-oil model in which the reservoir fluid is modeled by a pseudo-oil component and pseudo-gas component, compositional flow usually involves several to dozens of species. In reality, a hydrocarbon mixture may consist of tens to hundreds of substances, which is far more than the number of components in numerical simulations. 
% Considering reservoir fluids consist of tens to hundreds of components, the first idea that comes to mind is to lump the fluid mixture into a smaller number of pseudo-components \cite{Whitson2000} without losing much accuracy of the EOS model.
Thus, in order to accelerate phase equilibrium calculation, what first comes to mind is to lump the fluid mixture into a smaller number of pseudo-components \cite{Whitson2000} without losing much accuracy of the EOS model. Clearly, the less the number of components is, the more computational time can be saved and meanwhile the more accuracy will be lost. Despite this, the speedup caused by reducing the number of components is barely satisfactory, stimulating a large batch of researchers to develop acceleration strategies for phase equilibrium calculation over the last two decades. 
Among them, the reduction method was first proposed by Michelsen in 1986 \cite{Michelsen1986}, showing that only three equations need to be solved in two-phase flash problem under the Van der Waals mixing rule. The central idea of the reduced approach is to decrease the number of nonlinear equations and unknowns by taking advantage of the low-rank property of the binary interaction coefficient (BIC) matrix. However, his method, assuming all zero binary interaction coefficients, is severely circumscribed so that it's hardly applicable to real reservoir fluids. 
Later, Jesen and Fredenslund \cite{Jensen1987} extended the Michelsen’s research by introducing one non-hydrocarbon component for practical application. The nonzero BICs between the non-hydrocarbon and hydrocarbons yield five reduced variables, irrespective of the number of components, but still limits the applicability of their approach. 
A generalized reduction theorem was then presented by Hendriks \cite{Hendriks1988}, which gets rid of the restriction on the BIC matrix. He reformulated phase equilibria problems with the transformed chemical potential, as a function of reduction variables. By introducing the truncated spectral method, Hendriks and van Bergen \cite{Hendriks1992} simulated the two-phase equilibrium in the reduced space using Newton-Raphson iterations. 
Similarly, Pan and Firoozabadi proposed reduction models for phase stability testing \cite{Firoozabadi2000} and phase equilibrium calculation \cite{Pan2001} based on the spectral theory of linear algebra. For stability analysis, they found the tangent-plane distance (TPD) function in the reduced space has smoother surface accompanied with a unique minimum, both of which significantly improve the robustness of their method. On the other hand, by extensively testing the conventional approach and reduced approach of two-phase flash calculation, they concluded that the direct solution of the reduced model makes flash calculations more efficient and robust. 
Shortly afterwards, Li and Johns \cite{Li2006} developed their reduction model for rapid flash calculation by decomposing the BICs into two parts. The new phase split formulation includes at most six reduced parameters, either degenerating to the Michelsen's reduction model with three reduced parameters when all BICs are zero or the five-reduced-parameter model of Jensen and Fredenslund when the nonzero BICs only exist between a single component and the other components. Moreover, their method avoids solving the nonlinear Rachford-Rice equation \cite{Rachford1952}, a difficulty in phase splitting calculation, and meanwhile retains the accuracy of conventional approaches. 
Recently, Nichita and Graciaa \cite{Nichita2011} designed a reduced flash formulation with a new set of independent variables and error equations, leading to simpler expressions of the Jacobian matrix. Compared to the classical reduction methods, their approach exhibits better efficiency and meanwhile preserves the robustness. 
Instead of decomposing BIC matrices with the conventional spectral analysis approach, Gaganis and Varotsis \cite{Gaganis2013} introduced a new decomposition method to minimize the approximation errors of the energy parameter \cite{Gaganis2013}, which allows the phase equilibrium calculation to be performed at a given accuracy with fewer reduced variables. Even though all the above reduction methods have shown great potentials in accelerating flash calculation, recent researches \cite{Gorucu2015, Haugen2013, Michelsen2013, Petitfrere2015} questioned its efficiency against the conventional methods. It has been found that the reduced flash approach is faster only for mixtures with more than 15 components, making it more suitable for process simulation rather than compositional reservoir simulation where the number of components is usually less. 
% However, as mentioned in the beginning, a smaller acceleration than expected arises from the increasing nonlinearity with a large number of fluid components.
% This number is larger than the number of components often used in reservoir simulations, limiting their applicability. This behavior is attributed to an increase in the nonlinearity of the reformulated system of equations, leading to a smaller acceleration than expected \cite{ Micheslen2013}. 

% 3rd paragraph
In addition, there are many other approaches to speed up flash calculation within the framework of compositional simulator. Wang and Stenby \cite{Wang1994} developed a non-iterative flash calculation algorithm on the assumption that thermodynamic equilibrium is instantaneously established at any time of the simulation. With flash calculations decoupled from the compositional flow, the iterative flash procedure can be linearly approximated so that the mole numbers of components in each phase are computed based on the solution of compositional flow between the previous and current time step. There is no need to solve the Rachford-Rice equation for the phase saturation, thereby improving the efficiency of their method further. However, the reduction of CPU time is found to depends on a case by case. 
Given that repetitive phase stability analysis is time consuming, the shadow region method was proposed by Rasmussen et al. \cite{Rasmussen2003} in order to bypass unnecessary stability tests based on the location of any given condition in the phase diagram. It is assumed that the pressure, temperature or compositions have small changes in a certain grid block between two consecutive time steps. In the two-phase region, stability test procedure is completely bypassed by initializing flash calculation with the equilibrium solution in last time step. Furthermore, a one-sided stability analysis, initialized by the pre-stored non-trivial stability solution, replaces the full stability test within the shadow region, while outside the shadow region the distance to the critical point is used to determine whether stability analysis is skipped or not. Consequently, the computation time spent on flash calculation is significant reduced. 
On the other hand, to accelerate the phase splitting calculation, Voskov and Tchelepi \cite{Voskov2007,Voskov2008,Voskov2009} presented the Compositional Space Adaptive Tabulation (CSAT) method in which a tie-line table is constructed using the pre-stored phase split solutions to parameterize the compositional space. If the investigated composition is on the predetermined tielines or its extension, the phase equilibria solution can be approximated on the basis of the tieline compositions; otherwise, the standard flash procedure is performed and an additional tieline will be added into the lookup table. It is worth noting that the accuracy and efficiency of the CSAT strategy are heavily dependent on the number of stored tielines, as well as the tolerance set to accept the feed composition. As the tieline table becomes larger, we gain more accuracy but lose more efficiency as well. 
Inspired by Voskov and Tchelepi's research, Belkadi et al. \cite{Belkadi2011} suggested using the Tieline Distance-Based Approximation (TDBA) method to approximate flash solutions in the two-phase region. In their approach, the shortest distance from the new feed composition to the tieline in the same gridblock is used to determine if the rigorous flash solution is approximated, which minimizes the access and build-up process of the tieline table.
Recently, Yan et al. \cite{Yan2017} accelerated phase behavior calculations by interpolating initial approximations from the compositional space constructed, prior to the flash calculation, for reservoir fluids with both fixed and dramatically varying composition. For those hydrocarbon mixtures with significant compositional change during simulation, an additional pseudo key component is introduced, together with pressure, to control the phase behavior. Although their approach exhibits an enhanced performance under the Rachford-Rice preconditioning, the resolution of the compositional spaces constructed offline plays a critical role in such kind of method, as expected. 
Additionally, by taking advantage of the offline flash calculation, Wu et al. \cite{Wu2015} utilized the sparse grid surrogate model to approximate flash results during two-phase compositional simulations. Under a similar interpolation philosophy, a remarkable speedup of compositional simulation was achieved, but several intrinsic defects restrict the applicability of their method. Subsequently, Wu and Chen \cite{Wu2018} extended their work to approximate the equilibrium solution directly from the sparse grid instead of its surrogate model. The introduction of the "layer" concept and array structure makes the sparse grid construction efficient and inexpensive. However, two issues, including the violation of saturation boundedness and inaccurate detection of the single-phase region, need to be resolved for reliable application of sparse grid methods in phase equilibrium calculation.

% 4th paragraph
Apart from the aforementioned accelerating strategies, other alternatives, like negative flash calculation \cite{Li&Nghiem1982,Whitson1989} or linear transformation of the Rachford-Rice equation \cite{Li2007}, can be used to improve the efficiency of phase splitting calculations as well. All the above of speedup approaches is only the tip of iceberg and few of them can avoid iteratively solving phase equilibria problems. In addition to the iterative nature of flash algorithms, the increasing demand to high-resolution reservoir models, usually with millions of grid blocks, also sharply increases the computational costs of phase equilibrium calculations during compositional simulation. Thus, it is imperative to perform both phase stability testing and phase splitting calculation in a fast, non-iterative and robust way. With the rapid development of computing techniques, machine learning has become a powerful tool to deal with sophisticated and time-consuming problems. Since 2012, due to the breakthrough of AlexNet \cite{Krizhevsky2012}, deep learning, a subset of machine learning, made a revolutionary progress in image recognition, language processing, artificial intelligence, and so forth. The great potential of machine learning has drawn close attention from the petroleum industry, and remarkable efforts have been made in many aspects, such as reservoir characterization \cite{Korjani2016}, development planning optimization \cite{Ketineni2015} and permeability estimation \cite{ArayaPolo2018}. Furthermore, a number of researches have been devoted to speed up both phase stability test and phase splitting calculation using the machine learning method. 
Gaganis and Varotsis \cite{Gaganis2012} designed a stability discriminating function, which yields the same sign and zero points as the conventional tangent plane distance (TPD) approach, to eliminate the time-consuming iterations in stability analysis. Under the framework of Support Vector Machine (SVM) \cite{Burges1998,Cortes1995,Cristianini2000}, phase stability testing is treated as a binary classification problem and has an explicit expression on the basis of pressure, temperature and feed composition. For phase behavior calculation, they developed regression models using the artificial neural network (ANN) to predict the conventional equilibrium coefficients \cite{Gaganis2012EAGE} and later the prevailing reduced variables \cite{Gaganis2014}, in order to approximate the closed-form solution of the original flash formulation and the reduced one respectively. 
Instead of using SVM, Kashinath et al. \cite{Kashinath2018} utilized two relevance vector machine (RVM) classifiers for stability analysis. The supercritical classifier identifies the supercritical state at the given condition, while the subcritical classifier predicts the number of stable phases. Similar to Gaganis and Varotsis's approach, they estimated the equilibrium coefficients for the phase split problem using the regression model established on the basis of ANNs. 
In addition, Li et al. \cite{Tao2018} approximately solved the vapor-liquid equilibria problem by a deep learning model. Different from the aforementioned works, they directly estimated the equilibrium compositions rather than the equilibrium coefficients that work as the intermediate variables. A variety of effects are investigated, such as the number of hidden layers, the number of nodes per layer and the type of activation function, on the performance of the deep neural network, which helps them optimize their model to improve efficiency and accuracy. However, as a shortcoming of their work, they only discussed the cases where both vapor and liquid phases are present.

% 5th paragraph (what we want to do in this work)
In this study, instead of using the popular NPT-formulation, the phase equilibrium problem is formulated at constant volume and temperature due to the advantages of the NVT formulation. A dynamic model is designed for the isothermal isochoric phase equilibria problem and iteratively solved by Newton-Raphson methods with line search. To validate the capacity of our dynamic model, as well as the deep learning model, for handling complex fluid mixtures, three realistic reservoir fluids are investigated: one from Bakken shale formation includes 5 components and the other two from Eagle Ford shale formation consist of 8 and 14 components respectively. With the training data provided by the iterative flash calculation, we develop a deep neural network to approximate the NVT flash calculation and directly predict the equilibrium compositions of stable phases at any condition. To overcome the overfitting issue inherent in neural networks, the dropout technique is applied in combination of a regularized loss function. In addition, batch normalization significantly reduces the time spent on training the network model, thus enabling us to construct a more complex multi-level structure for better approximation. The network configuration is designed based on the results in \cite{Tao2018}, where the researchers investigated various effects on the performance of their deep neural network model. Different from the previous researches, one key effort of this study is to incorporate stability test and phase split calculation together so that both steps in the conventional framework of flash calculation can be accomplished by a single deep neural network. An independent variable is introduced in the training process to determine the number of stable phases. We present a number of numerical examples to validate the accuracy and efficiency of the proposed deep learning model. 

%\begin{itemize}
%	\item Since we mainly focus on phase equilibrium calculation itself in this work, the speedup approaches based on the result of flash calulation at previous time step or at adjacent grid cell, as well as those bypass methods, like the shadow region method, are not applied.
%	
%	\item Instead, the isothermal-isochoric phase equilibrium calculation is performed in parallel, with each phase split calculation initialized by the corresponding stability test. 
%	
%	\item The aformentioned parallel computation is accomplished by the MATLAB Parallel Computing Toolbox.
%\end{itemize}

% 6th paragraph (outlines of this paper)
The remainder of this paper is organized as follows. In Section \ref{sec: VTflash}, we formulate the dynamic model for the NVT flash problem in the bulk phase and compare our simulation results with the published ones. In Section \ref{sec: DLAlgorithm}, the deep learning algorithm is elaborated in detail with some techniques to enhance its performance. We discuss the prediction accuracy of the proposed deep neural network and compare its efficiency with the iterative flash calculation in Section \ref{sec: Results}. At the end, we make our conclusions in Section \ref{sec: Remarks}.
%The remainder of this paper is organized as follows. In Section \ref{sec: VTflash}, we formulate the dynamic model for the VT flash problem in the bulk phase and then compare our simulation results with the published ones. In Section \ref{sec: DLAlgorithm}, the deep learning method in acceleration of VT flash calculations is elaborated in detail. The performance (e.g. accuracy and e ciency) of the deep learning based flash calculation is discussed in Section \ref{sec: Results} by comparing with the standalone VT flash calculation. We make our conclusions in Section \ref{sec: Remarks}.

\section{NVT-flash calculations} \label{sec: VTflash}

Compared to the classical NPT-flash calculation, the NVT-flash exhibits some advantages and attracts a lot of attention from researchers in recent years. Without inverting the equation of state, the NVT-formulation has a unique solution and thus eliminate the root-selection procedure that usually takes place in NPT flash problems. Additionally, volume of pure substances under saturation pressure cannot be uniquely determined using the NPT formulation, since all states (two phases, single vapor or single liquid phase) of a pure substance share the same pressure on the phase boundary \cite{Jindrova2013}. Similar behavior has been observed in multicomponent mixtures with three or four phases as well \cite{Jindrova2015}. Despite the fact that the isothermal-isobaric phase equilibrium calculation is the most commonly used flash technique in compositional simulators, numerous efforts have been made to improve the computing performance of isothermal-isochoric phase equilibrium calculations and also extend its applications. Jindrov\'{a} and Miky\v{s}ka \cite{Jindrova2013} developed a fast and robust NVT-flash algorithm for two-phase equilibria problems by directly minimizing the total Helmholtz free energy. Subsequently, they extended their research to multiphase systems and investigated the equilibrium problem involving aqueous phase through the Cubic-Plus-Association EOS \cite{Jindrova2015}. To model the dynamic process from any non-equilibrium state to the equilibrium state, Kou and Sun \cite{Kou2016} established evolution equations for mole numbers and volume, which was solved by a well-designed energy-stable numerical algorithm. In addition to the aforementioned bulk phase flash problems, the confined phase behaviors at constant volume, temperature and moles have been studied by taking into account adsorption \cite{Cabral2005}, capillary pressure \cite{Kou2018,Li2018}, or confinement effect resulting from the interaction between fluid molecules and pore walls \cite{Luo2017, Travalloni2014}. In this study, we restrict our discussion to the bulk phase equilibrium of two-phase systems in which the fluid mixture is described by the Peng-Robinson EOS.

\subsection{The dynamic model for bulk phase equilibria}
Let us assume a fluid mixture of $M$ components stays in the vapor-liquid equilibrium at the given moles, volume and temperature. Thus, the total Helmholtz free energy of this system can be expressed as
\begin{equation} \label{eq: totalHelmholtz}
F = f\left(\mathbf{n}^{G}\right)V^{G} + f\left(\mathbf{n}^{L}\right)V^{L} \, ,
\end{equation}
where $\mathbf{n}^{\alpha} = \left[n_{1}^{\alpha}, \ldots, n_{M}^{\alpha}\right]^{T}$ and $V^{\alpha}$ denote the molar density and volume of phase $\alpha \left(\alpha = G, L\right)$, respectively. The Helmholtz free energy density  $f\left(\mathbf{n}\right)$ has the following form
\begin{equation} \label{eq: HelmholtzDensity}
f\left(\mathbf{n}\right) = RT\sum_{i=1}^{M}n_{i}\left(\ln n_{i}-1\right) - nRT\ln\left(1-bn\right) + \frac{a(T)n}{2\sqrt{2}b}\ln\left(\frac{1+\left(1-\sqrt{2}\right)bn}{1+\left(1+\sqrt{2}\right)bn}\right) \, .
\end{equation}
In eq. (\ref{eq: HelmholtzDensity}), $n =\displaystyle\sum_{j}n_{j}$ is the overall molar density, $R$ is the ideal gas constant and $T$ is the temperature. The Peng-Robinson parameters $a$ and $b$ are functions of the molar composition. Considering the total mole numbers $\mathbf{N}^{t} = \left[N_{1}^{t}, \ldots, N_{M}^{t}\right]^{T}$ and total volume $V^{t}$ are fixed, the following mole and volume constaints have to be satisfied whenever both vapor and liquid phases are present in the system
\begin{align}
	N_{i}^{G} + N_{i}^{L} &= N_{i}^{t} \, , \quad i = 1, \ldots, M \, , \\
	V^{G} + V^{L} &= V^{t} \, ,
\end{align}
where $N_{i}^{\alpha}$ represents the mole number of component $i$ in phase $\alpha$.

In any two-phase system, we can arbitrarily select the mole composition and volume of one phase as primary variables and compute the counterparts of the other phase by the aforementioned mole and volume constraints. Here the mole composition and volume of the vapor phase, denoted by $\mathbf{N}^{G}$ and $V^{G}$, are chosen as the primary variables so that the Helmholtz free energy is reduced as $F\left(\mathbf{N}^{G}, V^{G}\right)$. The partial derivatives of the Helmholtz free energy in eq. \ref{eq: totalHelmholtz} with respect to $N_{i}^{G}$ and $V^{G}$ yield
\begin{align}
	\frac{\partial F\left(\mathbf{N}^{G}, V^{G}\right)}{\partial N_{i}^{G}} &= \mu_{i}\left(\mathbf{n}^{G}\right) - \mu_{i}\left(\mathbf{n}^{L}\right) = \mu_{i}^{G} - \mu_{i}^{L} \, , \\
	\frac{\partial F\left(\mathbf{N}^{G}, V^{G}\right)}{\partial V^{G}} &= p\left(\mathbf{n}^{L}\right) - p\left(\mathbf{n}^{G}\right) = p^{L} - p^{G} \, ,
\end{align}
where $\mu_{i}^{\alpha}$ and $p^{\alpha}$ are the chemical potenial of component $i$ and pressure in phase $\alpha$, respectively. By applying the chain rule, the time derivative of the total Helmholtz free energy $F\left(\mathbf{N}^{G}, V^{G}\right)$ can be derived as follows
\begin{equation} \label{eq: TimeDerivative}
	\begin{split}
		\frac{\partial F}{\partial t} &= \sum_{i = 1}^{M}\frac{\partial F}{\partial N_{i}^{G}}\frac{\partial N_{i}^{G}}{\partial t} + \frac{\partial F}{\partial V^{G}}\frac{\partial V^{G}}{\partial t} \\
		&= \sum_{i = 1}^{M}\left(\mu_{i}^{G} - \mu_{i}^{L}\right)\frac{\partial N_{i}^{G}}{\partial t} + \left(p^{L} - p^{G}\right)\frac{\partial V^{G}}{\partial t} \, .
	\end{split}
\end{equation}

The mole and volume evolution equations are formulated based on Onsager's reciprocal principle, shown as below, to characterize the dynamic bulk phase equilibrium process
\begin{align} 
\frac{\partial N_{i}^{G}}{\partial t} &= \sum_{j=1}^{M} \psi_{i,j}\left(\mu_{j}^{G}-\mu_{j}^{L}\right) + \psi_{i,M+1}\left(p^{L}-p^{G}\right) \, , \label{eq: DynamicMole} \\
\frac{\partial V^{G}}{\partial t} &= \sum_{j=1}^{M} \psi_{M+1,j}\left(\mu_{j}^{G}-\mu_{j}^{L}\right) + \psi_{M+1,M+1}\left(p^{L}-p^{G}\right) \, . \label{eq: DynamicVolume}
\end{align}
According to the second law of thermodynamics, the total Helmholtz free energy in a closed system is supposed to dissipate with time, implying the Onsager coefficient matrix $\bm{\mathit{\Psi}} = \left(\psi_{i,j}\right)_{i,j=1}^{M+1}$ shall be negative definite. Following the similar strategy in \cite{Li2018}, we compute the Onsager coefficient matrix based on the initial equilibrium conditions. A revised modified Cholesky factorization is applied to ensure its negative definiteness, see \cite{Schnabel1999} for more details. It is worth mentioning that this well designed coefficient matrix helps to achieve the persistent dissipaition of the Helmholtz free energy, but also solve all evolution equations as a whole system in a simultaneous iteration approach. When the Helmholtz free energy is minimized, the investigated system reaches the equilibrium state, embodied in two aspects: the chemical equilibrium ($\mu_{i}^{G} = \mu_{i}^{L}$) and the pressure equilibrium ($p^{G} = p^{L}$).

\subsection{Numerical experiments}
\begin{figure}[H]
	\centering
%	\begin{subfigure}{0.6\textwidth}
%		\includegraphics[width=\textwidth]{Figures/VTFlash/PhaseEnvelope_Bakken_Bulk.pdf}
%		\caption{}
%	\end{subfigure}
%	\begin{subfigure}{0.6\textwidth}
%		\includegraphics[width=\textwidth]{Figures/VTFlash/PhaseEnvelope_EagleFord1_Bulk.pdf}
%		\caption{}
%	\end{subfigure}
%
%	\begin{subfigure}{0.48\textwidth}
%		\includegraphics[width=\textwidth]{Figures/VTFlash/PhaseEnvelope_EagleFord2_Bulk.pdf}
%		\caption{}
%	\end{subfigure}
	\hspace*{-1.5 cm}
	\includegraphics[width=0.6\textwidth]{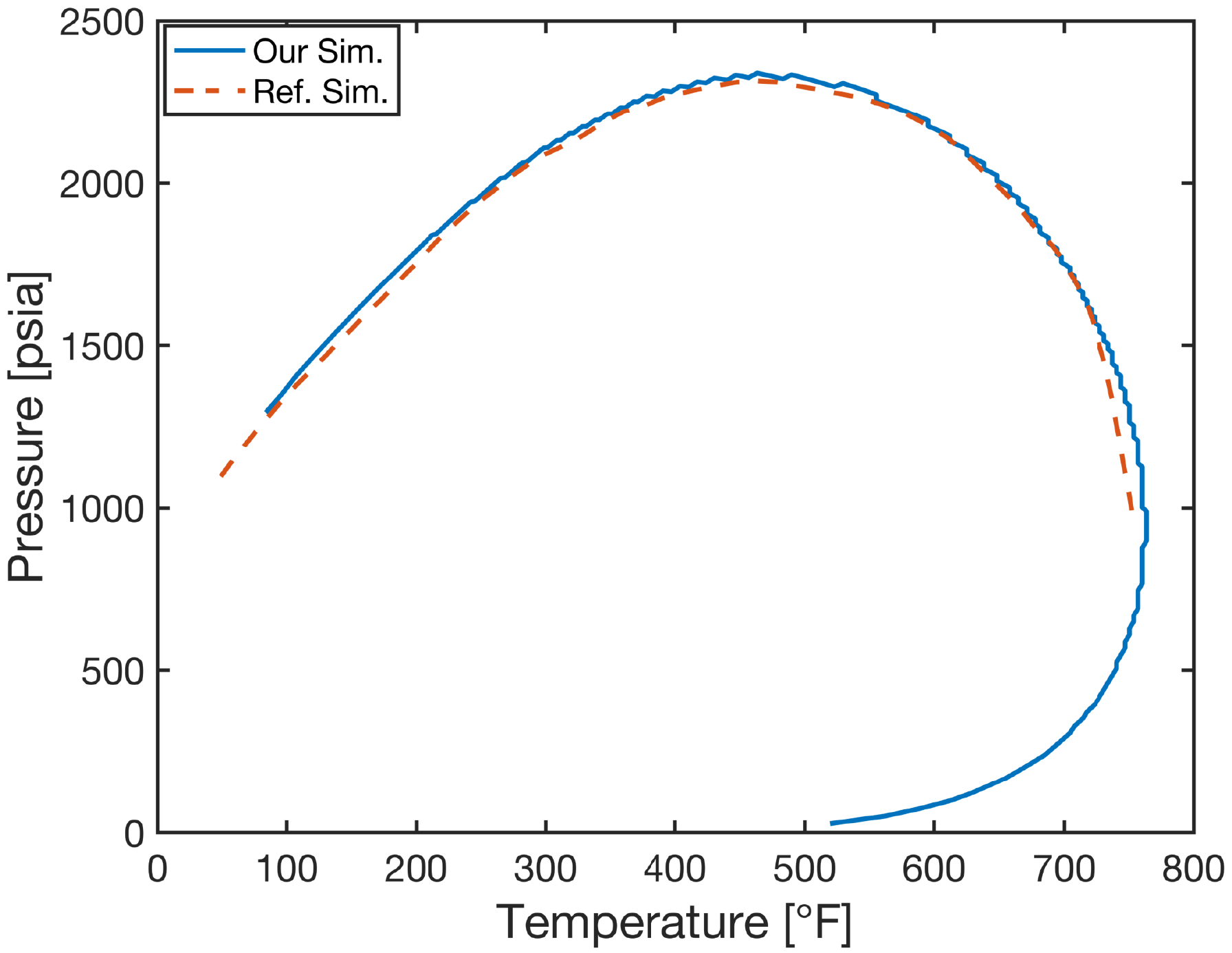}~
	\includegraphics[width=0.6\textwidth]{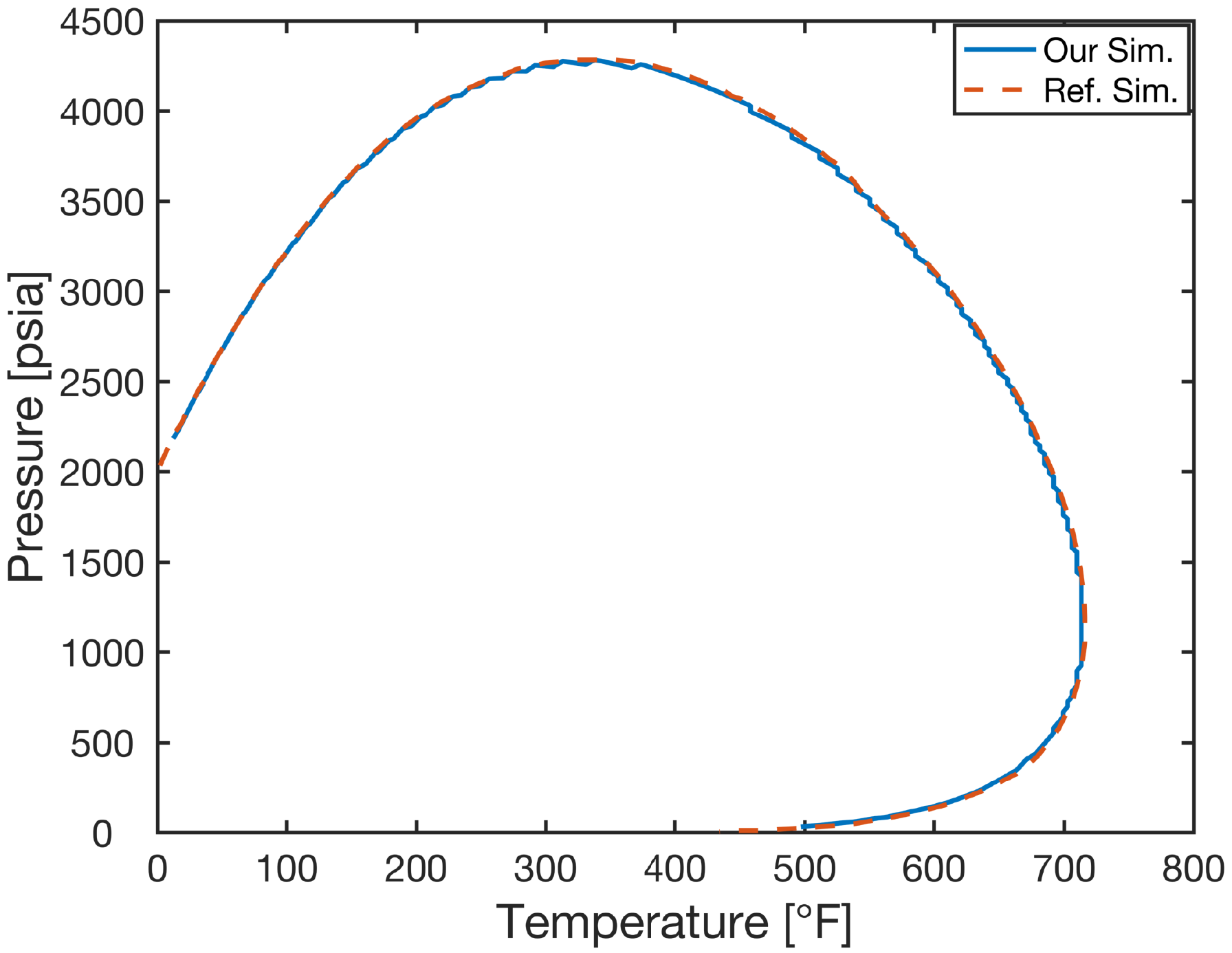} \\
	\includegraphics[width=0.6\textwidth]{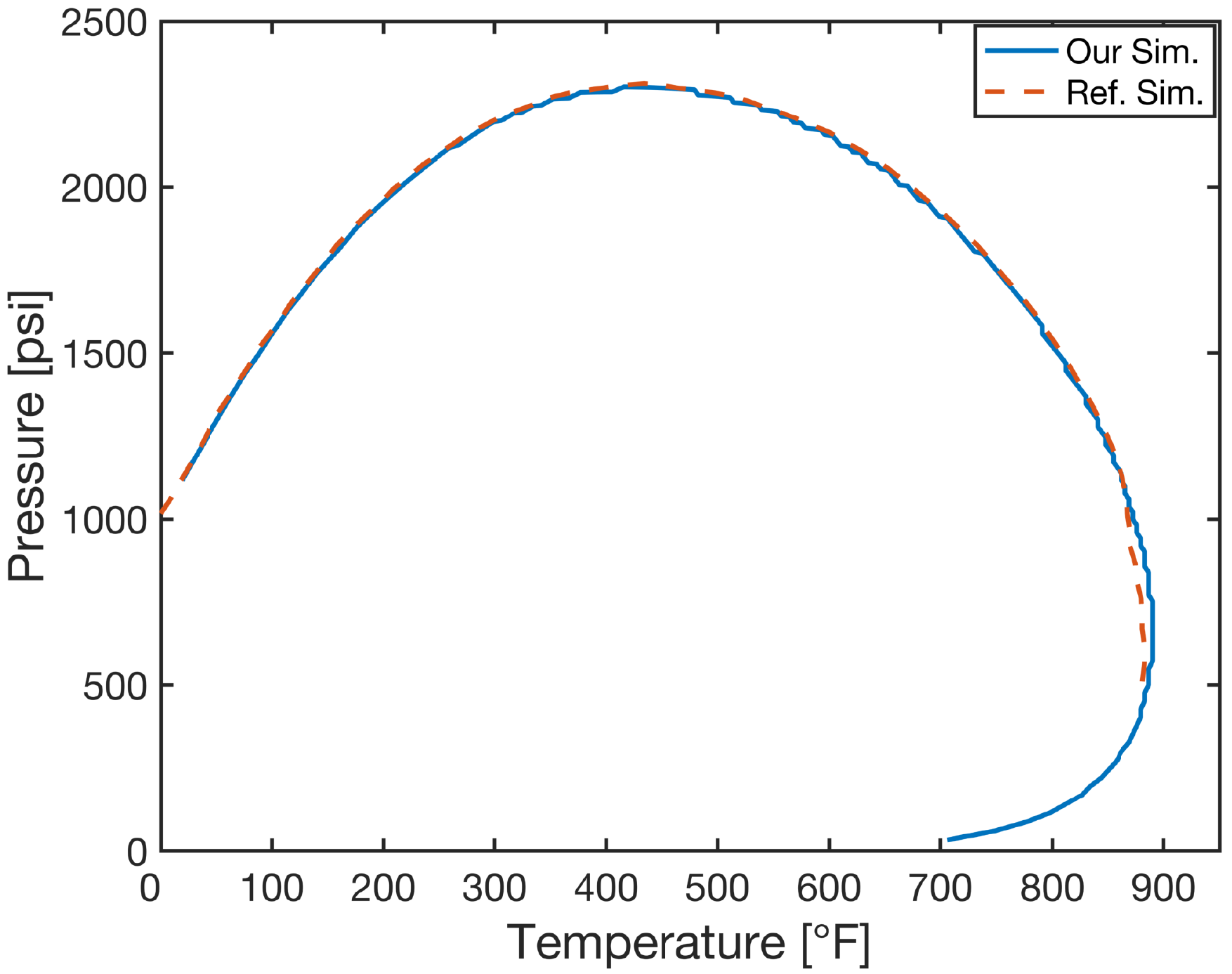}
	\caption{Bulk phase envelopes for: the Bakken oil (top left), the EagleFord1 oil (top right) and the EagleFord2 oil (bottom).}
	\label{fig: Figure1}
\end{figure}

To perform isothermal-isochoric phase equilibrium calculation, each phase splitting calculation is initialized by the stability test. Detailed implementation procedures of the bulk phase stability testing can refer to either the work of Miky\v{s}ka and Firoozabadi \cite{Mikyska2012} or the work of Kou and Sun \cite{Kou2018} with negligible capillary pressure. To construct an energy-stable time-marching scheme, which preserves the dissipation of the Helmholtz free energy, the convex splitting technique is employed to decompose the Helmholtz free energy density and chemical potential. In addition, we use a semi-implicit scheme \cite{Kou2018Flow} to discretize the mole and volume evolution equations. Specifically, the convex parts of the chemical potential and Helmholtz free energy density are treated implicitly while the concave parts are treated explicitly. The resultant nonlinear system is solved by the Newton-Raphson method with line search. To validate the accuracy of the proposed dynamic model, three reservoir fluid samples are tested: the Bakken oil consists of $5$ components and two Eagle Ford oils have $8$ and $14$ components. For brevity, the Eagle Ford oils are named by EagleFord1 for the eight-component sample and EagleFord2 for the other one with fourteen components. The molar composition and compositional properties for the Bakken, EagleFord1 and EagleFord2 oils are given in Table \ref{tab: Table1}--\ref{tab: Table3}. Accordingly, binary interaction coefficients for the three samples can be found in \cite{Zhang2016,Cui2018,Siripatrachai2016}. To achieve a smooth phase envelope, we uniformly discretize the computational domain of the overall molar density and temperature into a $201 \times 201$ grid. For the Bakken oil, the investigated molar density varies from $10$ to $10000 \; \text{mol}/\text{m}^{3}$ and the temperature ranges from $300$ to $850 \; \text{K}$. Additionally, the overall molar concentration of EagleFord1 and EagleFord2 samples are $\left[10, 12000\right]$ and $\left[10, 10000\right] \; \text{mol}/\text{m}^{3}$, while both of them have the same temperature interval $T \in \left[260, 850\right] \; \text{K}$. Figure \ref{fig: Figure1} shows the computed phase envelopes for the Bakken oil (top left), the EagleFord1 oil (top right) and the EagleFord2 oil (bottom) in sequence. It can be seen that the equilibrium solutions of the dynamic model agree with the published results in \cite{Zhang2016,Cui2018,Siripatrachai2016} very well. In the following, we use these NVT flash solutions, especially the equilibrium compositions of vapor and liquid phases, to train our deep learning model. 
% It is worth mentioning that the VT-flash calculations are parallelized by the MATLAB Parallel Computing Toolbox in order to reduce the computational time for data acquisition.

\begin{table}[H]
	\caption{Compositional parameters for the Bakken oil.} 
	\centering
	\label{tab: Table1}
	\begin{tabular}{c c c c c c}
		\hline
		Component & $z_{i}$ & $T_{c,i} \; (\text{K})$ & $P_{c,i} \; (\text{MPa})$ & $M_{w,i} \; (\text{mol}/\text{m}^{3})$ & $\omega_{i}$   \\
		\hline
		$\text{C}_{1}$ & 0.2506 &	190.606	& 4.6000  & 16.04 & 0.0080     \\
		$\text{C}_{2}$--$\text{C}_{4}$ & 0.2200 & 363.30 & 4.3100 & 42.82 & 0.1432    \\
		$\text{C}_{5}$--$\text{C}_{7}$ & 0.2000 & 511.56 & 3.4210  & 83.74 & 0.2474     \\
		$\text{C}_{8}$--$\text{C}_{9}$ & 0.1300 & 579.34 & 3.1320  & 105.91 & 0.2861     \\  
		$\text{C}_{10+}$ & 0.1994 & 788.74 & 2.1870 & 200.00 & 0.6869     \\     
		\hline
	\end{tabular}
\end{table}

\begin{table}[H]
	\caption{Compositional parameters for the EagleFord1 oil.} 
	\centering
	\label{tab: Table2}
	\begin{tabular}{c c c c c c}
		\hline
		Component & $z_{i}$ & $T_{c,i} \; (\text{K})$ & $P_{c,i} \; (\text{MPa})$ & $M_{w,i} \; (\text{mol}/\text{m}^{3})$ & $\omega_{i}$   \\
		\hline
		$\text{C}_{1}$ & 0.5816 & 190.56	& 4.5988  & 16.04 & 0.0110     \\
		$\text{C}_{2}$ & 0.0744 & 305.33 & 4.87180 & 30.07 & 0.0990    \\
		$\text{C}_{3}$ & 0.0417 & 369.83 & 4.2479  & 44.10 & 0.1520     \\
		$\text{nC}_{4}$ & 0.0259 & 418.71 & 3.7383  & 58.12 & 0.1948     \\  
		$\text{CO}_{2}$ & 0.0232 & 304.11 & 7.3739  & 44.01 & 0.2250     \\  
		$\text{C}_{5}$--$\text{C}_{6}$ & 0.0269 & 485.60 & 3.3767 & 76.502 & 0.2398 \\
		$\text{C}_{7+}$ & 0.1321 & 606.69 & 2.7083 & 122.96 & 0.3548     \\ 
		$\text{C}_{13+}$ & 0.0942 & 778.31 & 1.5598 & 255.28 & 0.7408     \\         
		\hline
	\end{tabular}
\end{table}

\begin{table}[H]
	\caption{Compositional parameters for the EagleFord2 oil.} 
	\centering
	\label{tab: Table3}
	\begin{tabular}{c c c c c c}
		\hline
		Component & $z_{i}$ & $T_{c,i} \; (\text{K})$ & $P_{c,i} \; (\text{MPa})$ & $M_{w,i} \; (\text{mol}/\text{m}^{3})$ & $\omega_{i}$   \\
		\hline
		$\text{C}_{1}$ & 0.31231 & 190.72	& 4.6409  & 16.04 & 0.0130     \\
		$\text{N}_{2}$ & 0.00073 & 126.22 & 3.3943  & 28.01 & 0.0400     \\
		$\text{C}_{2}$ & 0.04314 & 305.44 & 4.8842 & 30.07 & 0.0986    \\
		$\text{C}_{3}$ & 0.04148 & 369.89 & 4.2568  & 44.10 & 0.1524     \\
		$\text{CO}_{2}$ & 0.01282 & 304.22 & 7.3864  & 44.01 & 0.2250     \\  
		$\text{iC}_{4}$ & 0.01350 & 408.11 & 3.6480  & 58.12 & 0.1848     \\  
		$\text{nC}_{4}$ & 0.03382 & 425.22 & 3.7969  & 58.12 & 0.2010     \\  
		$\text{iC}_{5}$ & 0.01805 & 460.39 & 3.3336 & 72.15 & 0.2223 \\
		$\text{nC}_{5}$ & 0.02141 & 469.78 & 3.3750 & 72.15 & 0.2539 \\
		$\text{nC}_{6}$ & 0.04623& 507.89 & 3.0316 & 86.18 & 0.3007 \\
		$\text{C}_{7+}$ & 0.16297 & 589.17 & 2.7772 & 114.40 & 0.3739     \\ 
		$\text{C}_{11+}$ & 0.12004 & 679.78 & 2.1215 & 166.60 & 0.5260     \\     
		$\text{C}_{15+}$ & 0.10044 & 760.22 & 1.6644 & 230.10 & 0.6979     \\    
		$\text{C}_{20+}$ & 0.07306 & 896.78 & 1.0418 & 409.20 & 1.0456     \\      
		\hline
	\end{tabular}
\end{table}

\section{Deep learning algorithm} \label{sec: DLAlgorithm}

\begin{figure}[htbp]
	\centering
	\includegraphics[width = \linewidth]{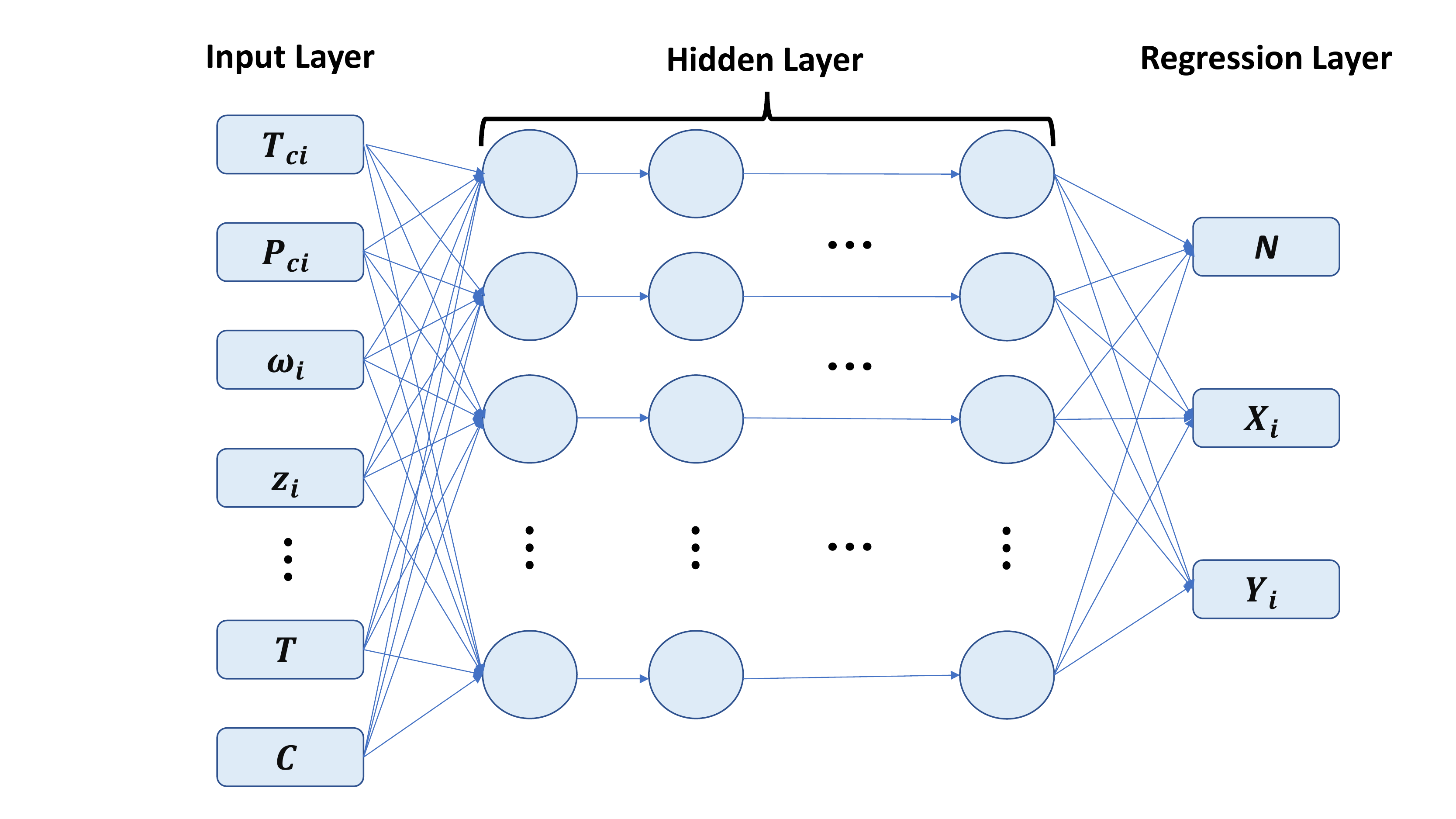}
	\caption{The deep neural network to approximate NVT flash calculation.} 
	\label{fig: ANNetwork}
\end{figure}

Artificial neural networks are computational models that are able to unearth the underlying correlations and features of input data and process them in a manner analogous to the biological nervous system. Such a characteristic enables the ANN model great problem-solving capacity. Various applications have been developed on the basis of ANNs, which are performed in either supervised or unsupervised approach, such as machine learning and pattern recognition. Unlike the unsupervised ANN model, the supervised machine learning tries to seek a function that can be used to predict the values of desired variables from input data under a given accuracy. In this study, instead of using the common machine learning technique, a deep neural network, which consists of multiple nonlinear transformation layers, is applied to replace the iterative NVT-flash algorithm in order to accelerate phase equilibrium calculations. Figure \ref{fig: ANNetwork} displays the schematic diagram of the proposed deep neural network, which processes the input data, including compositional properties (e.g. critical pressure ($P_{c}$), critical temperature ($T_{c}$), acentric factor ($\omega$)), overall mole fraction ($z$), the equilibrated vapor/liquid molar composition given by the NVT flash calculation, in a sequential approach. Each activation layer transforms the output of the preceding layer and yields results as input to the following one. This hierarchical structure enables the deep neural network model to extract useful features and patterns layer by layer and then learn complicated correlations behind the enormous data volume. Benefiting from the powerful capacity of the deep learning model, we accomplish both stability test and phase split calculation at once by a single network model so that the number of phases is determined free of additional stability analysis. This differs from the recent researches on the acceleration of flash calculation by machine learning methods, in which the stability test is preserved to determine if the investigated fluid is stable or not as the prevailing flash framework does. As a result, the output of the regression layer in the proposed network model includes three parts: the number of phases ($N$), the equilibrium mole fraction of vapor components $X_{i}$ and the equilibrium mole fraction of liquid components $Y_{i}$ at the specified overall molar density ($C$) and temperature ($T$).  
% The activation function of this layer is fixed as linear. 

The performance of a neural network model can be evaluated based on the loss, which is defined as the difference between the model output and the ground truth. Here we use the mean square error as the loss function, evaluated in each iteration, to describe the deviations of the model prediction from the solution of iterative flash calculation. However, overfitting, as the inherent issue of neural networks, gravely affects the performance of the deep learning algorithm, which manifests that the trained model exhibits excellent performance based on the training data while yields terrible predictions for its further application. This inconsistent behavior is attributed to the numerous parameters of the model waiting for optimization and insufficient training data, thus making the model over-parameterized. In order to protect the deep neural network model from the overfitting issue, additional constraints are imposed on the parameters to reduce the freedom of the model. Generally, if the model is overfitted, the norm of the weight parameters will become very large. For the avoidance of the overfitting problem, an additional constraint is added on the norm of weight parameters to penalize the large weights. In practice, we introduce a regularization term, related to the norm of the weights, in the loss function, which yields
% The original loss function for deep learning, which is the mean squared loss in our problem, can be formulated as
%\begin{equation} \label{eq: LossFun}
%L = \frac{1}{N} \sum_{n = 1}^{N} {\Vert \textbf{o} - \hat{\textbf{o}} \Vert}^2,
%\end{equation}
%where $N$ is the number of the training data, $\bf{o}$ is the output of the model and $\hat{\bf{o}}$ is the observation value. After adding the L2 weight decay term, the loss function becomes:
\begin{equation} \label{eq: LossFun}
L = \frac{1}{N}\sum_{n = 1}^{N} {\Vert \textbf{o} - \hat{\textbf{o}} \Vert}^2 + \lambda {\Vert \bf{W} \Vert}^2_{2} 
\end{equation}
where $N$ is the number of the training data, $\bf{o}$ is the model output, $\hat{\bf{o}}$ is the observation value, $\bf{W}$ is the weight parameters, and $\lambda$ is the regularization coefficient for weight penalization. In addition to the regularized loss function, we also the dropout technique to resolve the overfitting problem, which reduces the freedom of the deep neural network model by abandoning certain layers during the training stage. 
% For example, if we apply the dropout technique to a certain layer with the keep probability as $p \, (0<p<1)$, then, during each training stage, each node of that layer would first be evaluated independently with the probability of $p$ being kept or the probability of $1-p$ being discarded. 
In particular, with a specified threshold probability $p \, (0<p<1)$, each node of a certain layer is evaluated independently and retained if its probability is less than the threshold value; otherwise, the node will be removed. If any node on the investigated activation layer vanishes, all the nodes and connections will be abandoned, yielding a reduced neural network for training. After this, the full model is recovered by inserting the removed layer back and enters the next training cycle.

\begin{figure}[htbp]
	\centering
	\includegraphics[width = \linewidth]{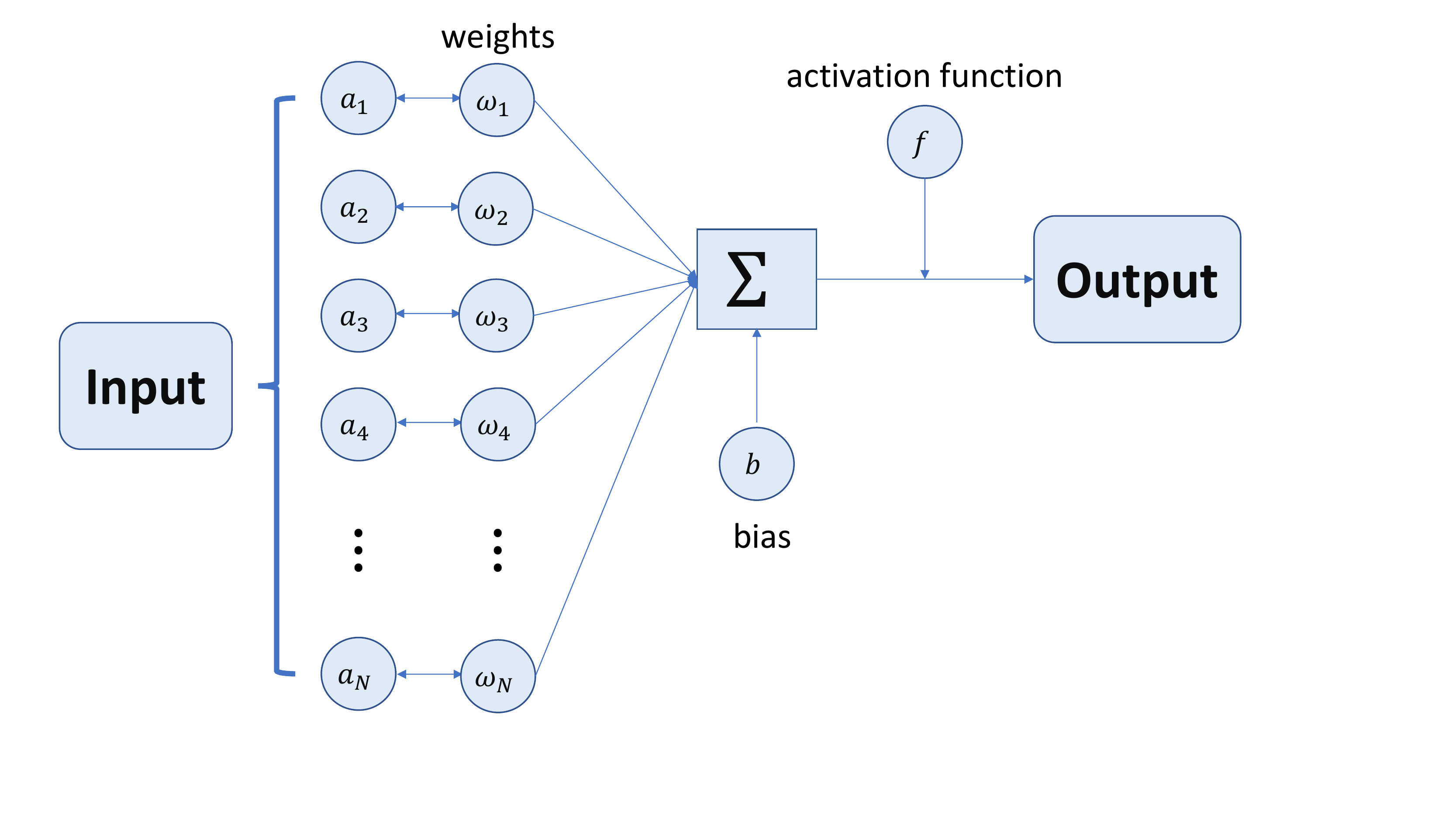}
	\caption{The flow chart of the learning process in each node.} 
	\label{fig: FlowChart}
\end{figure}

% The initialization of the neural network model is of vital importance, which can affect the convergence speed and even the final model's performance. If the weights are initialized with very small values, the variance of the input signal vanishes across different layers and eventually drops to a very low value, which reduces the model complexity and may hurt the model's performance. If the weights are initialized with very large values, the variance of the input signal tends to increase rapidly across different layers. That may cause gradient vanishing or explosion, which increases the difficulty of training a working model. Since we usually initialize the weights with a Gaussian distribution, to control the variance of the signal, it is desirable to initialize the weights with a variance $\delta$ to make the variance of the output of a layer the same as that of the input of the layer. 
In order to design a robust and efficient neural network, it is crucial to properly initialize the weight parameters since they have significant influence on the convergence rate and performance of the deep learning algorithm. If the initial weights are underestimated, the variance of the input data will drop to a very small value through some intermediate transformations, which reduces the model complexity and damage its performance. On the other hand, the overestimated weight parameters increase the variance of the input rapidly, making the gradient either vanishing or exploding and eventually failing to train the model. In this study, the weight parameters are initialized following the Gaussian distribution. To control the variance during the training process, we then initialize the weights with a variance $\delta$ and make the output variance always same to the input variance on the same activation layer. For example, in Figure \ref{fig: FlowChart}, the variance of $y$ is expected to expand as follows
%\begin{align}
%var(y) & = var(w_1*a_1+w_2*a_2+...+w_n*a_n+b) \nonumber \\
%& = var(w_1)*var(a_1) + var(w_2)*var(a_2) +...+ var(w_n)*var(a_n) \nonumber \\
%& \stackrel{(1)}{=} n*var(w_i)*var(a_i), 
%\end{align}
\begin{equation}
	\begin{split}
		\mathrm{var(y)} & = \mathrm{var}\left(w_{1}\times a_{1}+w_{2}\times a_{2} + \ldots + w_{n}\times a_{n} + b \right) \\
		& = \mathrm{var}(w_{1})\times \mathrm{var}(a_{1}) + \mathrm{var}(w_{2})\times \mathrm{var}(a_{2}) + \ldots + \mathrm{var}(w_{n})\times \mathrm{var}(a_{n}) \\
		% & \stackrel{(1)}{=} n*var(w_i)*var(a_i), 
		& = n\times \mathrm{var}(w_{i})\times \mathrm{var}(a_{i})
	\end{split}
\end{equation}
with the identical distribution assumption for all $w_{i}$ and $a_{i}$. Since we want $ \mathrm{var(y)} $ equal to $ \mathrm{var}(a_{i})$, the following equality has to be satisfied
\begin{equation}
n\times \mathrm{v}ar(w_{i}) = 1 \, .
\end{equation}
Therefore, the weights of each layer are initialized following the Gaussian distribution with the variance of $1/n$, where $n$ is the number of weights in that layer. This initializer is known as the Xavier initializer.

% Training deep learning model is notoriously time-consuming, because of the large number of parameters belonging to different layers. Not only is the optimization for such a large number of parameters internally time-consuming, but there are some undesirable properties of the multi-layer model which makes the convergence process slow. One property of the deep learning method is that the distribution of each layer's input might change because the parameters of the previous layer are usually changed during training, which is usually referred to as ``internal covariate shift''. To solve the problem, batch normalization is proposed. In addition to normalize the original input of the model, which is the input of the first layer, this technique makes the normalization part of the model and performs normalization on hidden layers for each training batch during the training stage. Batch normalization enables larger learning rates and can accelerate the convergence speed by 10 times.
It has to be mentioned that training a deep neural network is time-consuming due to massive parameters waiting for optimization over all the hidden layers. In addition, some undesirable properties of the multi-level model, as well as the input data of different scale at each activation layer, slow down the training process as well. Among others, the varying distribution of the activations play a critical role in slow convergence of training deep neural networks, since each layer has to adjust itself to a new distribution in each training step.
% But the problem appears in the intermediate layers because the distribution of the activations is constantly changing during training. This slows down the training process because each layer must learn to adapt themselves to a new distribution in every training step. This problem is known as internal covariate shift.
To speed up the learning process, batch normalization is employed to adjust and scale the outcomes of the preceding layer, which can be regarded as a preprocessing step for each activation layer. Consequently, a significant improvement of the training speed is achieved, which enables to train more complicated models in a reasonable amount of time.

In this study, we develop a deep neural network model to approximate the iterative NVT-flash calculation using TFlearn on a MacBook Pro laptop with Intel Core i7 processor. Admittedly, training neural networks takes a large amount of time. 
% The time spent on training iterations depends on the data size. Generally, training our neural network model will be completed in less than 10 minutes if the results of VT flash calculation on a $151 \times 151$ grid are provided as the input. 
However, the trained model can be repeatedly used instead of iteratively solving nonlinear flash problems, together with the offline training process in advance, giving the deep learning model great potential to speed up phase equilibrium calculation.
A simplified flowchart of the working process in each node of the proposed model is presented in Figure \ref{fig: FlowChart}.
Let $\mathbf{a}_{i}$ and $\mathbf{y}_{i}$ denote the input and output of the $i$th activation layer, respectively. The output of the $i$th layer is given by
\begin{equation} \label{eq: LayerOutput}
\mathbf{y}_{i} = \textit{f}_{i}(\mathbf{W}_{i}\times\mathbf{a}_{i} + \mathbf{b}_{i}) \, ,
\end{equation}
where $\mathbf{W}_{i}$ is the weight, $\mathbf{b}_{i}$ is the bias, and $\textit{f}_{i}$ is the activation function of the $i$th layer. Considering the output of a given layer is the input of its next one, the network model in Figure \ref{fig: ANNetwork}, for instance, with three activation layers, can be expressed as follows, 
\begin{equation} \label{eq: NetworkModel}
\mathbf{o} = \textit{f}_{3}(\mathbf{W}_{3}\times\textit{f}_{2}(\mathbf{W}_{2}\times\textit{f}_{1}(\mathbf{W}_{1}\times\mathbf{x}_{1}+\mathbf{b}_{1})+\mathbf{b}_{2})+\mathbf{b}_{3}) \, .
\end{equation}
As shown in eq. (\ref{eq: NetworkModel}), the activation function is the source of the nonlinearity in neural networks, which helps the deep learning algorithm approximate complex functions and gives it great problem-solving capability. There are a variety of activation functions, including Rectified linear unit (ReLU), Parameteric rectified linear unit (PReLU), TanH, Sigmoid, Softplus, Softsign, Leaky rectified linear unit (Leaky ReLU), Exponential linear unit (ELU) and Scaled exponential linear unit (SELU). Among them, the most commonly used activation functions are ReLU 
\begin{equation}
	f(x) = 
	\begin{cases}
	0,        & \text{if } x<0\\
	x,      & \text{if } x \geq 0 \, ,
	\end{cases}
\end{equation}
and Sigmoid
\begin{equation}
	\sigma(x) = \frac{1}{1+\exp(-x)} \, .
\end{equation}

\section{Results and Discussion} \label{sec: Results}
As mentioned in Section \ref{sec: VTflash}, the training data for the deep learning model are provided by solving the dynamical model of NVT flash problems. Three real reservoir fluids are investigated, including the five-component Bakken oil, eight-component EagleFord1 oil and fourteen-component EagleFord2 oil. Detailed compositional parameters for each reservoir fluid are given in Table \ref{tab: Table1}--\ref{tab: Table3}. Following the theory and practice in Section \ref{sec: DLAlgorithm}, a deep neural network is designed with 5 activation layers, each of which contains 100 nodes, and totally 4000 iterations. Additionally, "ReLU" is chosen as the activation function. It is worth mentioning that the performance of this network configuration has been validated in \cite{Tao2018}. To investigate the effect of data size on the performance of the deep neural network, we calculate NVT flash results for the EagleFord1 oil on the same computational domain with size of $51 \times 51$, $71 \times 71$, $101 \times 101$, $151 \times 151$, $201 \times 201$ and $301 \times 301$ uniform grids. In addition, the results for the Bakken oil and EagleFord2 oil are computed on the specified concentration and temperature intervals, which are uniformly divided into $301\times301$ grids. All the eight data sets are used to train the proposed neural network, and the efficiency and accuracy of the trained model are tested. One key effort of this study is to investigate the possibility in achievement of both stability testing and phase splitting calculation by a single network model, which is different from the conventional two-step framework that all the preceding researches on acceleration of phase equilibrium calculations by machine learning follow.
% One key effort of this study is to investigate the possibility of incorporating the stability test with phase splitting calculation, which is processed using our deep learning approach to test its capability to handle this problem.

\subsection{Deep Learning Model Training} \label{subsec: Loss}
As shown in Figure \ref{fig: ANNetwork}, the compositional properties of fluid components, as well as the temperature and overall concentration are used as the input, and the deep neural network predicts mole fractions of components in both vapor and liquid phases. The key parameters of the model are the weights of each activation layer, which control the model prediction under the given input data. At the beginning, those weights are initialized randomly, implying that the model will yield useless results initially. To approximate the NVT flash calculation by the deep learning model, we optimize the weight parameters to fit the equilibrium vapor and liquid molar composition iteratively solved from NVT flash problems. In the following, $90\%$ of the data are used to train our network model, while the remaining $10\%$ data are used for validation. We denote by $D_{\text{train}}$, $D_{\text{test}}$ the number of training and testing data, respectively. 

Table \ref{tab: Table4} presents the training time ($t_{\text{train}}$) and testing time ($t_{\text{test}}$) of the deep neural network with different data sets. Here $t_{\text{test}}$ represents the time that the trained model spent on equilibrium predictions for the same data size of flash problems. For instance, it takes $7.9$ seconds for the trained model to predict the mole fractions of EagleFord1 oil on a $201 \times 201$ grid. In addition, the mean absolute error ($\varepsilon_{a}$) and relative error ($\varepsilon_{r}$) are computed with respect to iterative flash solutions and presented in Table \ref{tab: Table4} as well.  Clearly, for the EagleFord1 oil, as the number of input data becomes larger, the training time significantly increases, while the testing time doesn't change too much. Furthermore, we observe that both absolute and relative prediction errors continue to decrease with the data volume increasing. It is worth noting that the predicted accuracy has an unremarkable improvement when the data size of the EagleFord1 oil increases from $201 \times 201$ to $301 \times 301$. This weak accuracy enhancement is not worthy of the effort that takes more than double CPU time. With the same data volume, it seems the more components are involved, the larger prediction error the deep neural network model exhibits. However, the Bakken oil sample makes an exception and yields greater error than the Eagle Ford oils. This might be attributed to the underneath correlations between different components differ from the investigated fluid mixtures so that the trained network model may yield different accuracy. Essentially, the composition of the Bakken oil is quite different from the compositions of two Eagle Ford samples, which exhibit some similarities to some extent. This may explain our observation disagrees with the expectation.

\begin{table}[H]
	\caption{Performance of the deep learning model for different data samples.} 
	\centering
	\label{tab: Table4}
	\resizebox{\linewidth}{!}{
		\begin{tabular}{ c c c c c c c c }
			\hline 
			source & $N_{\text{train}}$ & $N_{\text{test}}$ & $t_{\text{train}} \; \left(s\right)$ & $t_{\text{test}} \; \left(s\right)$ & $\varepsilon_{a}$ & $\varepsilon_{r}$ \\
			\hline 
			% Bakken$\left(301 \times 301\right)$ & 81540 & 9061 & 2352 & 9.2 & 0.02362526 & 0.02885216 \\
			Bakken$\left(301 \times 301\right)$ & 81540 & 9061 & 2352 & 9.2 & 0.02362 & 0.02885 \\
			% EagleFord1$\left(51 \times 51\right)$ & 2340 & 261 & 42 & 4.1 & 0.02946343 & 0.03581562 \\
			EagleFord1$\left(51 \times 51\right)$ & 2340 & 261 & 42 & 4.1 & 0.02946 & 0.03582 \\
			% EagleFord1$\left(71 \times 71\right)$ & 4536 & 505 & 134 & 5.9 & 0.01929609 & 0.02410446 \\
			EagleFord1$\left(71 \times 71\right)$ & 4536 & 505 & 134 & 5.9 & 0.01930 & 0.02410 \\
			% EagleFord1$\left(101 \times 101\right)$ & 9180 & 1021 & 235 & 6.3 & 0.01377607 & 0.01723054 \\
			EagleFord1$\left(101 \times 101\right)$ & 9180 & 1021 & 235 & 6.3 & 0.01378 & 0.01723 \\
			% EagleFord1$\left(151 \times 151\right)$ & 20520 & 2281 & 512 & 7.1 & 0.01300997 & 0.01646667 \\
			EagleFord1$\left(151 \times 151\right)$ & 20520 & 2281 & 512 & 7.1 & 0.01301 & 0.01647 \\
			% EagleFord1$\left(201 \times 201\right)$ & 36361 & 4040 & 1232 & 7.9 & 0.01260957 & 0.01580003 \\
			EagleFord1$\left(201 \times 201\right)$ & 36361 & 4040 & 1232 & 7.9 & 0.01261 & 0.01580 \\
			% EagleFord1$\left(301 \times 301\right)$ & 81540 & 9061 & 2531 & 8.6 & 0.01252074 & 0.01571335 \\
			EagleFord1$\left(301 \times 301\right)$ & 81540 & 9061 & 2531 & 8.6 & 0.01252 & 0.01571 \\
			% EagleFord2$\left(301 \times 301\right)$ & 81540 & 9061 & 2468 & 8.9 & 0.01684125 & 0.01963720 \\
			EagleFord2$\left(301 \times 301\right)$ & 81540 & 9061 & 2468 & 8.9 & 0.01684 & 0.01964 \\
			\hline 
		\end{tabular}
	}
\end{table}

\begin{table}[H]
	\caption{Comparison of the computational time of iterative flash calculation to the training time and testing time of the deep neural network for the EagleFord1 oil with different data size.} 
	\centering
	\label{tab: Table5}
	\resizebox{0.7\linewidth}{!}{
		\begin{tabular}{ c c c c }
			\hline 
			source & $t_{\text{flash}} \; \left(s\right)$ & $t_{\text{train}} \; \left(s\right)$ & $t_{\text{test}} \; \left(s\right)$ \\
			\hline 
			EagleFord1$\left(51 \times 51\right)$ & 61 & 42 & 4.1 \\
			EagleFord1$\left(71 \times 71\right)$ & 122 & 134 & 5.9 \\
			EagleFord1$\left(101 \times 101\right)$ & 237 & 235 & 6.3 \\
			EagleFord1$\left(151 \times 151\right)$ & 547 & 512 & 7.1\\
			EagleFord1$\left(201 \times 201\right)$ & 1021& 1232 & 7.9 \\
			EagleFord1$\left(301 \times 301\right)$ & 2100 & 2531 & 8.6\\
			\hline 
		\end{tabular}
	}
\end{table}

Table \ref{tab: Table5} compares the computational time of NVT flash calculations to the training time and testing time of the deep neural network for the EagleFord1 oil with different data size. At low data volume, the time spent on doing iterative flash calculations and on training the network model is very close. On the other hand, the training time is clearly more than the computational time of iterative flash calculations when the data size reaches to $201 \times 201$ or higher. This may result from the fact that the deep neural network not only processes the data but also dig out the correlations behind them, the latter of which is supposed to consume a large amount of time to approximate the complicated nonlinear functions. At this point, it is believed the training time of a deep neural network is comparable to the computational time of iterative flash calculations under the same data volume. Also, we observe that the testing time is much less than the computational time of flash calculations. When the data size reaches to $301\times301$, the trained model makes predictions 244 times faster than the iterative flash calculation. It has to be mentioned that training the network model is a one-time process for a given flash problem. Overall, the deep learning approach will save plenty of time for phase equilibrium predictions.

\begin{figure}[H]
	\centering
	\includegraphics[width = \linewidth]{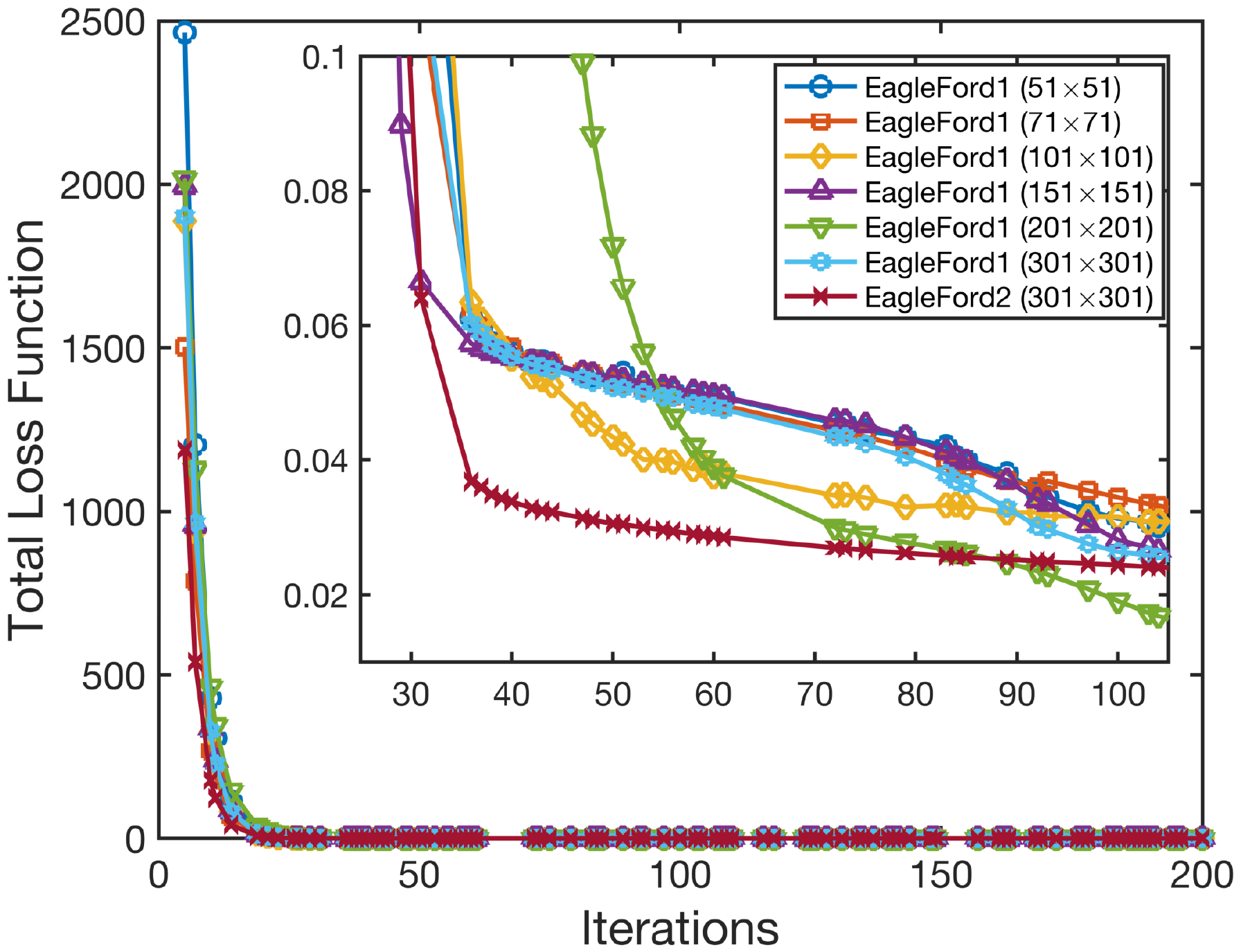}
	\caption{The variation of the total loss function for EagleFord1 and EagleFord2 oils. The inset figure zooms in the first 100 iterations with the loss function less than $0.1$.} 
	\label{fig: LossFun}
\end{figure}

Figure \ref{fig: LossFun} shows the loss function decreases with iterations, indicating the predictions of the trained mode are approaching to the results of NVT flash calculation. The inset figure displays the local variation of the total loss function after a sharp decrease within the first $100$ iterations, which is indistinguishable in the main figure. Clearly, the difference between deep learning predictions and iterative flash solutions continues to decrease, but the decline rate significantly slows down after about $40$ or $50$ iterations. For the EagleFord1 oil, the increasing data size, from $51 \times 51$ to $201 \times 201$, helps to accelerate the reduction of the loss function. However, such a speedup effect vanishes with more data used to train the model. In comparison, the loss function of the EagleFord2 oil exhibits a faster decline rate with the same data size (e.g. $301\times301$) to the EagleFord1 oil until it reaches a certain threshold, below which the loss function remains almost steady.

\subsection{Phase Number Characterization} \label{subsec: Number}
\begin{figure}[H]
	\centering
	\includegraphics[width = \linewidth]{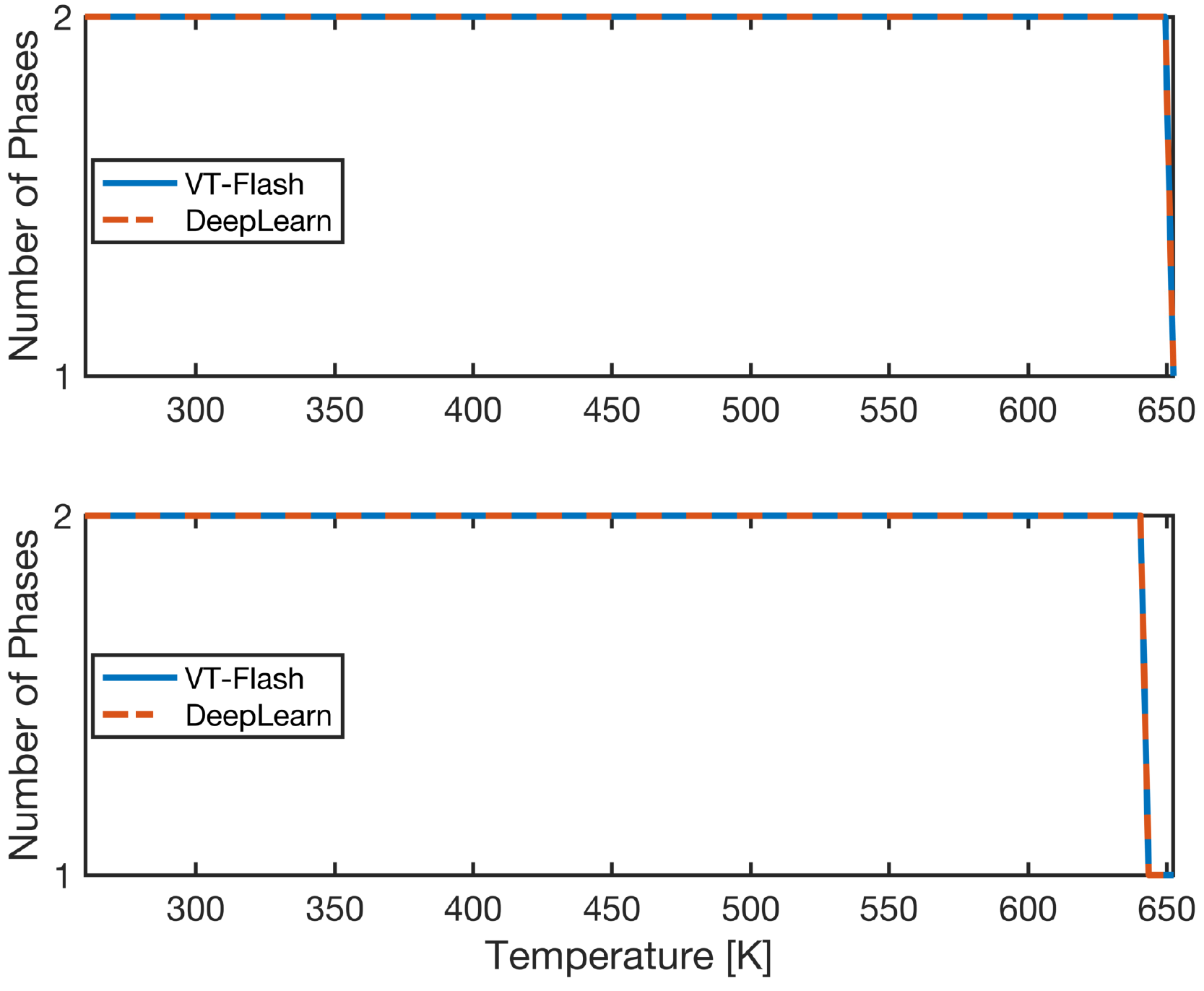}
	\caption{Number of phases, predicted by the deep learning model (dash line) and iterative flash calculation (solid line), as a function of temperature under the specified overall concentration $1988 \; \text{mol}/\text{m}^{3}$ (top) and $2648 \; \text{mol}/\text{m}^{3}$ (bottom), repsectively.} 
	\label{fig: PhaseNum}
\end{figure}

Previous researches follow the conventional framework of flash calculation, which consists of two separate stages: stability test and phase split calculation. In contrast, one key effort in this study is training the deep neural network to automatically determine the number of phases without additional stability analysis. Under this guideline, the proposed neural network model achieves both stability testing and phase splitting calculation simultaneously, and predicts the number of phases together with the molar compositions of vapor and liquid phases, as shown in Figure \ref{fig: ANNetwork}. To validate the prediction accuracy of our model on phase stability problems, we compare the number of phases computed by the deep learning model and iterative flash calculation. Figure \ref{fig: PhaseNum} displays the phase number of the EagleFord1 oil as a function of the temperature at two overall concentrations $1988 \; \text{mol}/\text{m}^{3}$ (top) and $2648 \; \text{mol}/\text{m}^{3}$ (bottom), respectively. Note that we only plot the results under the subcritical region in Figure \ref{fig: PhaseNum}. For the EagleFord1 oil, when the temperature exceeds $650 \; \text{K}$, it enters the supercritical state. Overall, our deep learning model successfully captures the phase transition process from the vapor-liquid region to the single liquid region. The single-phase state, either single vapor or single liquid, is determined based on the fact that the mole fraction of each component in the existing phase should approach to its overall mole fraction. It also can be seen that the predictions of the deep learning model near the phase boundary agree with iterative flash solutions very well in these two cases. In addition, we observe the phase number predicted by the deep learning model slightly deviates from iterative flash solutions near the boundary between the supercritical region and subcritical region. Particularly, the deep neural network predicts the EagleFord1 oil enters the supercritical state ahead of one condition in Figure \ref{fig: PhaseNum} (bottom).

\subsection{Preidiction Evaluation} \label{subsec: Comparison}
\begin{figure}[H]
	\centering
	\includegraphics[width = \linewidth]{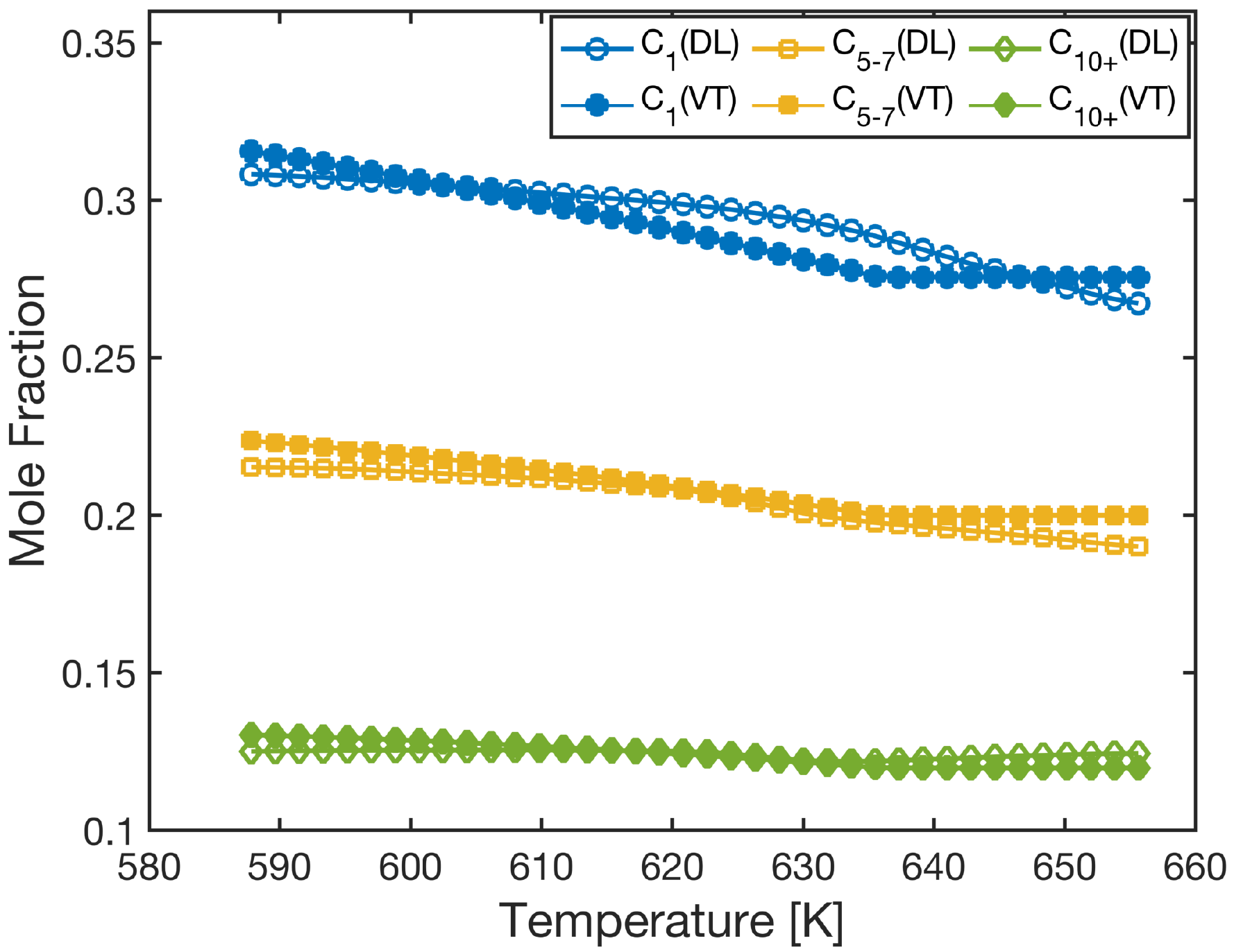}
	\caption{Mole fractions of $\text{C}_{1}$, $\text{C}_{5-7}$ and $\text{C}_{10+}$ in the vapor phase of Bakken oil predicted by the deep learning approach (open symbols) and iterative flash calculation (solid symbols).} 
	\label{fig: Bakken_Vapor}
\end{figure}

\begin{figure}[H]
	\vspace*{-2.5 cm}
	\centering
	\includegraphics[width = \linewidth]{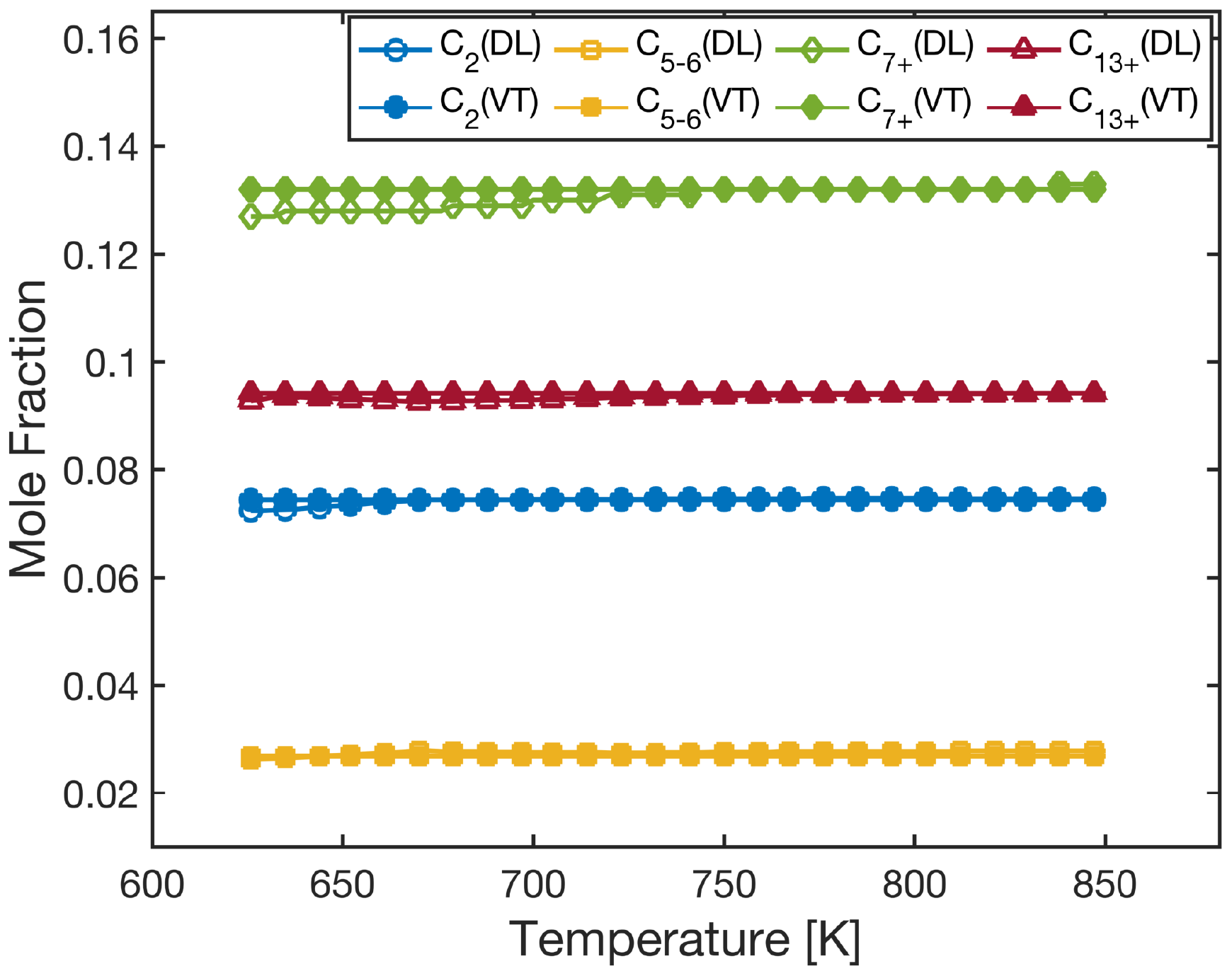}
	\caption{Mole fractions of $\text{C}_{2}$, $\text{C}_{5-6}$, $\text{C}_{7+}$ and $\text{C}_{13+}$ in the single-phase EagleFord1 oil. The open and solid symbols represent the deep learning prediction and iterative flash solution, respectively.} 
	\label{fig: EagleFord1_Vapor}
\end{figure}

\begin{figure}[H]
	\centering
	\includegraphics[width = \linewidth]{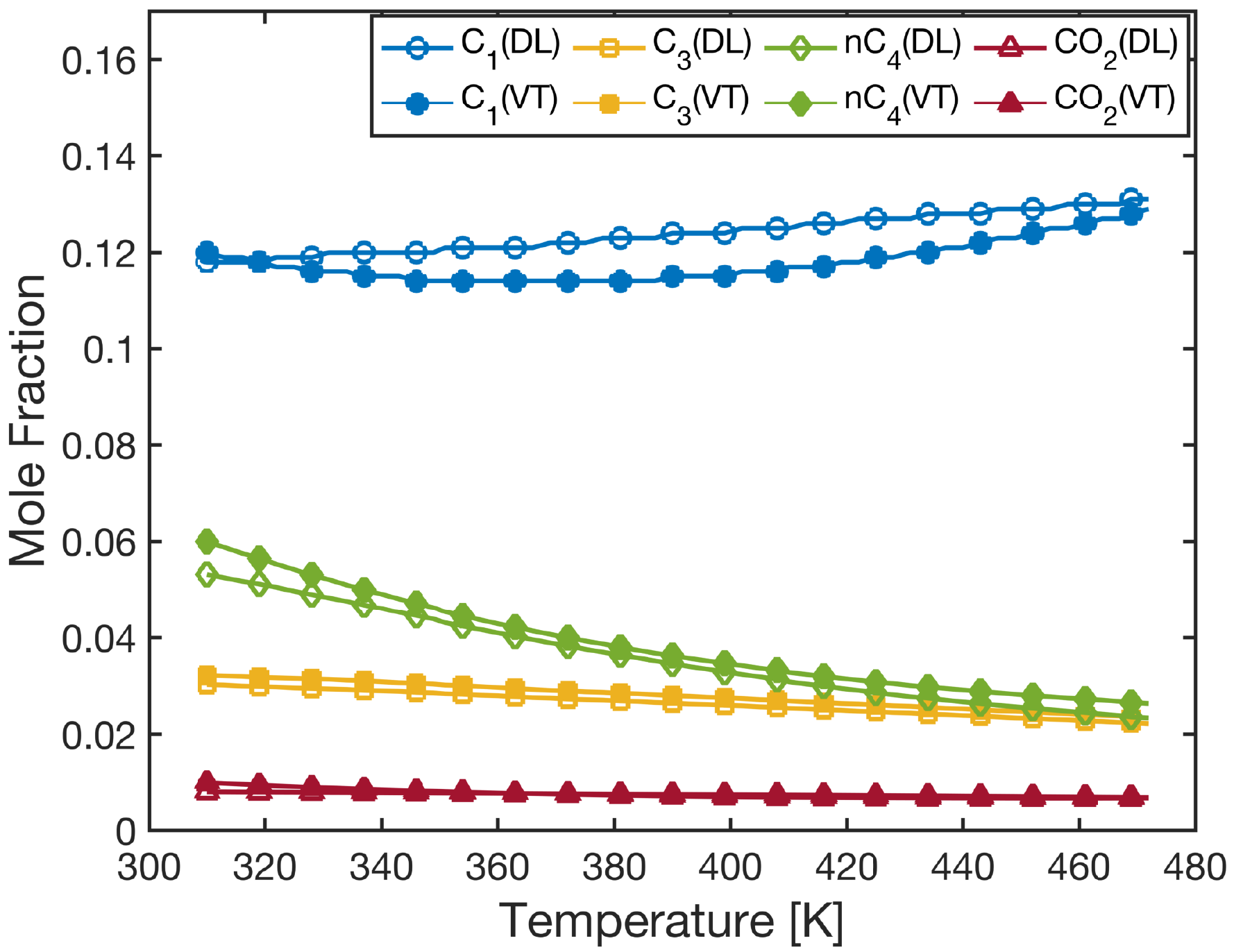}
	\caption{Mole fractions of $\text{C}_{1}$, $\text{C}_{3}$, $\text{nC}_{4}$ and $\text{CO}_{2}$ in the liquid phase of the EagleFord1 oil. The open and solid symbols represent the deep learning prediction and iterative flash solution, respectively.} 
	\label{fig: EagleFord1_Liquid}
\end{figure}

In Section \ref{subsec: Loss}, the accuracy of our deep learning model shows as the small prediction errors in Table \ref{tab: Table4}. Here we visually compare molar compositions computed by the deep learning and iterative flash approaches to further illustrate this accuracy. Figure \ref{fig: Bakken_Vapor} displays the mole fractions of $\text{C}_{1}$, $\text{C}_{5-7}$ and $\text{C}_{10+}$ in the vapor phase of the Bakken oil as a function of the temperature at $C = 343 \; \text{mol}/\text{m}^{3}$. The open and solid symbol represent the deep learning prediction and iterative flash solution, respectively. Even though the mole fraction profiles computed by both approaches exhibit some opposite variation tendency, the deep neural network is considered to yield good predictions under certain confidence, especially for the heaviest component $\text{C}_{10+}$. 
Figure \ref{fig: EagleFord1_Vapor} shows the mole fractions of $\text{C}_{2}$, $\text{C}_{5-6}$, $\text{C}_{7+}$ and $\text{C}_{13+}$ remain constant at the specified temperature interval, implying the EagleFord1 oil locates in the single-phase region at the given overall concentration. It can be seen the deep learning predictions agree with the iterative flash solutions very well. Figure \ref{fig: EagleFord1_Liquid} presents the mole fractions of $\text{C}_{1}$, $\text{C}_{3}$, $\text{nC}_{4}$ and $\text{CO}_{2}$ in the liquid phase of the EagleFord1 oil at another temperature interval when $C = 1810 \; \text{mol}/\text{m}^{3}$. The mole fraction of $\text{C}_{1}$ predicted by the network model deviates from the results of NVT flash calculation more distinctly than the other light components. Overall, the optimized deep neural network exhibits a good performance on the prediction accuracy. Admittedly, in some cases there are some slight differences.

\begin{figure}[H]
	\centering
	\includegraphics[width = \linewidth]{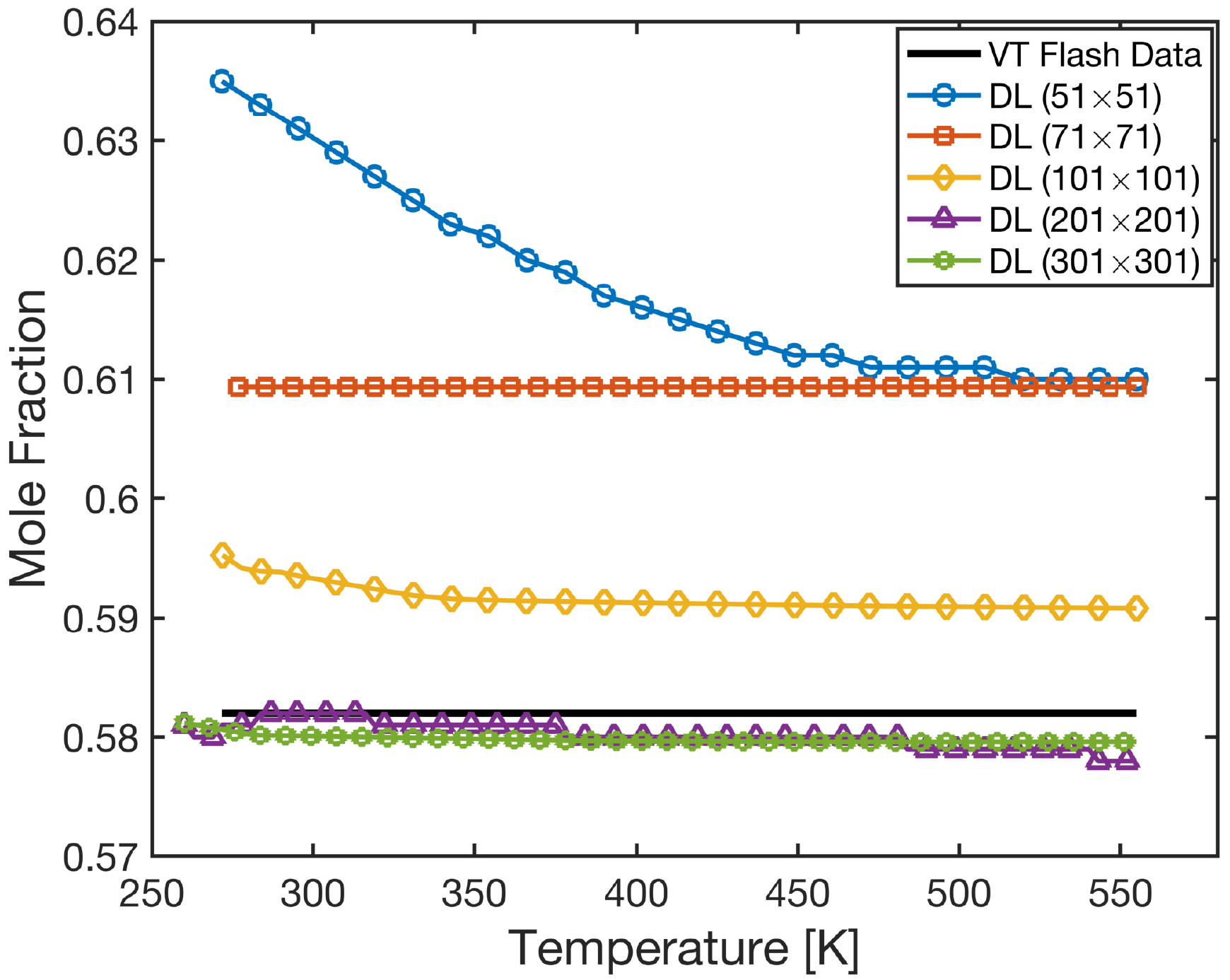}
	\caption{Mole fraction of $\text{C}_{1}$ predicted by the deep learning model under different data size when the EagleFord1 oil is at the single-phase state. The black line represents the solutions of iterative flash calculation.} 
	\label{fig: EagleFord1_Single_C1}
\end{figure}

The effect of data size on the prediction accuracy is validated in Figure \ref{fig: EagleFord1_Single_C1}. With the temperature increasing, the mole fraction of $\text{C}_{1}$ keeps constant at its overall mole fraction $0.5816$, denoted by the black line, which means the EagleFord1 oil stays in the single-phase region under the given conditions. As expected, the less data we use to train the model, the more deviation takes place in the outcomes of the deep neural network. 
% It is easy to find that the estimation accuracy is improved significantly if we increase the data source size from $51 \times 51$ to $201 \times 201$. 
It is found that the deep neural network with the data size of $201 \times 201$ gives pretty good approximations, which exhibit some slight fluctuations in comparison to the steady predictions computed based on the $301 \times 301$ data size. When the data size increases from $201 \times 201$ to $301 \times 301$, the insignificant accuracy improvement, shown in Figure \ref{fig: EagleFord1_Single_C1},  is not worthy of doubling the computational time as we see in Table \ref{tab: Table4}. Thus, a moderate number of data should be used to train the model in order to balance the accuracy and efficiency of the deep learning model.

\section{Conclusions} \label{sec: Remarks}

In this study, we establish a deep neural network to approximate NVT flash calculation so as to speed up phase equilibrium calculations. A dynamic model, comprising the mole and volume evolution equations, is iteratively solved to prepare data for training the proposed network model. Three real reservoir fluids, including a five-component Bakken oil and two Eagle Ford oils with eight and fourteen components, are investigated, showing that both the dynamic model of NVT flash problems and deep learning model can handle complex fluid mixtures. To increase the prediction accuracy, we reformulate the loss function and employ the dropout technique to reduce the overfitting problem inherent in neural networks. Furthermore, batch normalization is applied to accelerate the training process of the deep learning model. It has to be mentioned that the network configuration is designed based on our previous research, which studied various effects on the performance of deep neural networks for flash calculation approximation. Different from the previous researches, our deep learning model gets rid of the constraint of the conventional flash framework where stability test precedes phase split calculation, so that it can determine the number of phases without additional stability analysis. In addition, the proposed network model is able to identify the single-phase state, either single vapor or single liquid. A variety of examples are presented to exhibit the accuracy and efficiency of the deep neural network model. We find that training a deep learning model becomes much more expensive above a threshold data volume, however, with less enhanced accuracy. Although the training time is comparable to the computational time of iterative flash calculations, the trained model yields much faster predictions, at most 244 times for the investigated cases, under large data quantity and meanwhile preserve good accuracy. Moreover, the trained model can be repeatedly used for the given flash problem after a one-time offline training process, which gives the deep learning model great potential to speed up the time-consuming compositional simulation as well. It is worth mentioning that this study paves the path for our future research on acceleration of phase equilibrium calculations. In addition to improving the algorithm for higher prediction accuracy, some important effects in unconventional reservoirs, including capillary pressure and nanopore confinement, will be taken into account in the future work to validate the capability of deep neural networks in confined phase equilibrium problems.

\section*{Acknowledgement}
The authors greatly thank for the Research Funding from King Abdullah University of Science and Technology (KAUST) through the grant BAS/1/1351- 01-01 and the support from the National Natural Science Foundation of China (No. 51874262).

\end{document}